# Seizure intervention via diffusion mechanisms

Erik D. Fagerholm[1,*], Rosalyn J. Moran[1], Robert Leech[1]

[1] Centre for Neuroimaging Sciences, Department of Neuroimaging, IoPPN, King's College London

* Corresponding author: erik.fagerholm@kcl.ac.uk

We begin by demonstrating that the neuronal state equation from Dynamic Causal Modelling takes on the form of the discretized Fokker-Planck equation upon the inclusion of local activity gradients within a network. Using the Jacobian of this modified neuronal state equation, we show that an initially unstable system can be rendered stable via a decrease in diffusivity. Therefore, if a node within the system is known to be unstable due to a tendency toward epileptic seizures for example, then our results indicate that intervention techniques should take the counter-intuitive approach of mirroring activity in the nodes immediately surrounding a pathological region – effectively fighting seizures with seizures. Finally, using a two-region system as an example, we show that one can also tune diffusivity in such a way as to allow for the suppression of oscillatory activity during epileptic seizures.

## Structural-dynamic causal modelling

Prior to Dynamic Causal Modelling (DCM)[1], causal models in neuroimaging considered the evolution of states $x$ as a static function of themselves:

$$x = f(x, v, \theta), \quad [1]$$

where $f$ is a nonlinear function; $v$ are external inputs; and $\theta$ are model parameters.

DCM then extended [1] by considering the rate of change of the states with respect to time:

$$\frac{dx}{dt} = f(x, v, \theta), \quad [2]$$

which reduces to [1] in the limiting case that inputs $v$ vary slowly relative to the states $x$.

We can take a further step in the progression from [1] to [2] by allowing states to vary not just with respect to time, but with respect to any arbitrary number of dimensions $\mu$, such that:

$$\sum_j \frac{\partial x}{\partial \mu_j} = f(x, v, \theta). \quad [3]$$

We then retain time as the first dimension ($\mu_1 = t$) and assign all additional terms to describe the rate of change of states along the structural dimensions of the network.

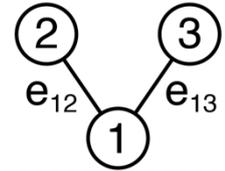

For example, if we consider the simple three-node network on the right, we describe the evolution of the first node using [3] as follows:

$$\sum_{j=1}^{3} \frac{\partial x_1}{\partial \mu_j} = \frac{dx_1}{dt} + \sigma_1 \left( \frac{\partial x_1}{\partial e_{12}} + \frac{\partial x_1}{\partial e_{13}} \right) = f(x_1, v, \theta), \quad [4]$$

where $\sigma_1$ is a constant of proportionality with dimensions of $[\sigma_1] = T^{-1}$; and e.g. $\frac{\partial x_1}{\partial e_{12}}$ describes the rate of change of the first node's activity along the edge connecting the first and second nodes.

We write the general form of [4] as follows for the $i^{th}$ node:

$$\frac{dx_i}{dt} + \sigma_i \sum_j^N k_{ij} \frac{\partial x_i}{\partial e_{ij}} = f(x_i, v, \theta), \qquad [5]$$

where $N$ is the total number of nodes in the network; and $k_{ij}$ is the adjacency matrix element connecting the $i^{th}$ and $j^{th}$ regions – ensuring that only local structural gradients are taken into account.

The structural gradients $\frac{\partial x_i}{\partial e_{ij}}$ in [5] can be discretized by expressing these in terms of the difference in state values (at a given time) at the $i^{th}$ and $j^{th}$ nodes, such that:

$$\dot{x}_i = f(x_i, v, \theta) + \sigma_i \sum_j^N k_{ij}(x_i - x_j), \qquad [6]$$

where we see that in the limiting case that structural gradients are zero $(x_i = x_j)$, the second term on the left-hand side is zero and [6] reduces to the DCM neuronal state equation in [2].

We therefore obtain a third level of neuronal state equations beginning with causal models in [1], followed by DCMs in [2], and ending with structural-DCMs in [6] – with each level reducing to the former within appropriate limits.

**Structural DCM and the Fokker-Planck equation**

The second term on the right-hand side of [6] can be re-written as follows:

$$\sum_j^N k_{ij}(x_i - x_j) = Nx_i - \sum_j^N k_{ij}x_j = \sum_j^N (\delta_{ij}d_i - k_{ij})x_j = \sum_j^N l_{ij}x_j, \qquad [7]$$

where $d_i$ is the degree of the $i^{th}$ region; and $l_{ij}$ is the graph Laplacian matrix element connecting the $i^{th}$ and $j^{th}$ regions.

Using [7] we can then write [6] as follows:

$$\dot{x}_i = f(x_i, v, \theta) + \sigma_i L x_i,  \quad [8]$$

and as the graph Laplacian is defined as the discretized version of the Laplace operator $\nabla^2$ the second term on the right-hand side describes a diffusion process, hence lending an interpretation to $\sigma$ as a diffusion coefficient.

It can be seen that the state equation in [8] takes the form of the discretized Fokker-Planck equation, with both a drift $f(x_i, v, \theta)$ and a diffusion $\sigma_i L x_i$ term. Therefore, if [8] is shown to model neural timeseries then, as the Fokker-Planck equation is a statement of the conservation of probability, this implies a conservation of neuronal firing or depolarisation.

**Linear stability**

In order to model neural timeseries, we first linearize [6]:

$$\dot{x}_i = \sum_j^N p_{ij} x_j + \sigma_i \sum_j^N k_{ij}(x_i - x_j) + q_{mi} v, \quad [9]$$

where $p_{ij}$ reduces to the intrinsic DCM coupling matrix $a_{ij}$ in the limiting case of zero structural gradients $(x_i = x_j)$; and $q_{mi}$ describes the effect of the $m^{th}$ external input on the $i^{th}$ region, which reduces to the extrinsic DCM coupling matrix $c_{mi}$ in the limiting case of zero structural gradients.

As the extrinsic coupling $q_{mi}$ does not affect a system's resilience to perturbation[2], the stability of a system described by [9] is characterised by the following Jacobian:

$$J = \begin{bmatrix} p_{11} + \sigma_1 \sum_{j \neq 1}^N k_{1j} & \cdots & p_{1N} - \sigma_1 k_{1N} \\ \vdots & \ddots & \vdots \\ p_{N1} - \sigma_N k_{N1} & \cdots & p_{NN} + \sigma_N \sum_{j \neq N}^N k_{Nj} \end{bmatrix} \quad [10]$$

**Seizure intervention type I: increasing stability**

We begin by making the following four assumptions: 1) A neural system that is prone to epileptic seizures can be described by [9], for which all parameters can be recovered via Bayesian model inversion; 2) We know where the pathological region lies; 3) Seizure onset is due in part to dynamic instability in the system, as characterised by the sum of eigenvalues of [10] having a positive Real component; 4) Epileptic seizures can be mitigated by increasing dynamic stability.

We proceed via a similar logic to that used by Turing to derive the conditions that allow for an initially stable system to develop spatial patterns via an instability caused by diffusion mechanisms[3]. However, we begin with the opposite premise and ask the following question: can we push the initially unstable (prone to seizures) system described by [10] into a stable regime by altering the diffusive properties in the system?

To answer this, let us first assume that the first region is known to be prone to seizures. We then transform the diffusion coefficient associated with the first region as follows:

$$\sigma_1 \rightarrow \alpha\sigma_1 \qquad [11]$$

where $\alpha$ is a constant.

The system described by [10] is initially unstable, such that the sum of its eigenvalues (given by the trace) has a positive Real component:

$$trJ = p_{11} + \sigma_1 \sum_{j \neq 1}^{N} k_{1j} + \cdots + p_{NN} + \sigma_N \sum_{j \neq N}^{N} k_{Nj} > 0. \qquad [12]$$

We then transform [10] via [11] to obtain:

$$J' = \begin{bmatrix} p_{11} + \alpha\sigma_1 \sum_{j \neq 1}^{N} k_{1j} & \cdots & p_{1N} - \alpha\sigma_1 k_{1N} \\ \vdots & \ddots & \vdots \\ p_{N1} - \sigma_N k_{N1} & \cdots & p_{NN} + \sigma_N \sum_{j \neq N}^{N} k_{Nj} \end{bmatrix}, \qquad [13]$$

which has been rendered stable, such that its trace is now negative:

$$trJ' = p_{11} + \alpha\sigma_1 \sum_{j \neq 1}^{N} k_{1j} + \cdots + p_{NN} + \sigma_N \sum_{j \neq N}^{N} k_{Nj} < 0, \qquad [14]$$

which, together with [12], says that:

$$trJ' = trJ + \sigma_1(\alpha - 1) \sum_{j \neq 1}^{N} k_{1j} < 0, \qquad [15]$$

where we know from [12] that $trJ > 0$ and by definition the diffusion coefficient $\sigma_1 > 0$, so that:

$$(\alpha - 1) \sum_{j \neq 1}^{N} k_{1j} < 0 \quad \Rightarrow \quad \alpha < 1 \qquad [16]$$

i.e. a necessary but not sufficient condition for an initially unstable system described by [10] to be rendered stable via a change in diffusivity [11], is that this change must act in a way as to decrease diffusivity [16].

It is not possible to change a diffusion coefficient as in [11], as this is an intrinsic property of a system. However, if we look at the diffusion term $\sigma_i \sum_j^N k_{ij}(x_i - x_j)$ in the governing equation of motion [9], we see that in lieu of decreasing $\sigma_i$ we can instead decrease the gradient $x_i - x_j$. This can be achieved by targeting the external driving input $v$ via the extrinsic coupling $q_{mi}$ in such a way as to decrease the gradients between a given pathological region and its neighbouring nodes. Therefore, we take the counter-intuitive measure of effectively mirroring the activity of a region that is itself prone to excess activity.

**Seizure intervention type II: disrupting periodicity**

Here we proceed on the assumption that seizures can be suppressed by pushing a neural system's dynamics from oscillatory (Imaginary eigenvalues) to non-oscillatory (purely Real eigenvalues) activity.

As a simple example, we consider a two-region system, for which [10] becomes:

$$J = \begin{bmatrix} p_{11} + \sigma_1 k_{12} & p_{12} - \sigma_1 k_{12} \\ p_{21} - \sigma_2 k_{21} & p_{22} + \sigma_2 k_{21} \end{bmatrix}, \quad [17]$$

which we assume to be initially in an oscillatory regime, such that the radicand in the eigenvalue $\lambda$ equation satisfying $det(J - \lambda I) = 0$ is negative:

$$(trJ)^2 - 4detJ < 0 \quad [18]$$

Upon transforming the diffusion coefficient as in [11], equation [17] becomes:

$$J' = \begin{bmatrix} p_{11} + \alpha\sigma_1 k_{12} & p_{12} - \alpha\sigma_1 k_{12} \\ p_{21} - \sigma_2 k_{21} & p_{22} + \sigma_2 k_{21} \end{bmatrix}, \quad [19]$$

which has now been rendered non-oscillatory, such that:

$$(trJ')^2 - 4detJ' > 0, \quad [20]$$

which, together with [18] tells us that:

$$\alpha < \frac{2}{k_{12}\sigma_1}(p_{11} - p_{22} - 2p_{21} + \sigma_2 k_{21}) - 1, \quad [21]$$

i.e. a system described by [17] that is initially in an oscillatory regime can be rendered non-oscillatory by transforming diffusivity [11], as long as the necessary but not sufficient condition in [21] is satisfied. Note that [21] only holds for a two-region system, but similar expressions can be calculated on a case-by-case basis by examining the radicands in the eigenvalue equation satisfying $det(J - \lambda I) = 0$ for a given network.